# Fluctuating Immersed Material (FIMAT) Dynamics for Fully Resolved Simulation of the Brownian Motion of Particles

Yong Chen, Neelesh A. Patankar[*]

Department of Mechanical Engineering, Northwestern University, Evanston, IL 60208-3111

**Abstract**

Fluctuating hydrodynamics based techniques have been developed in recent years for the simulation of Brownian motion of particles. These "mesoscale" simulation tools are viable approaches for problems where molecular dynamics simulations may be deemed expensive. We have developed a rigid constraint–based formulation where the key idea is to assume that the entire domain is a fluctuating fluid. Rigid motion constraints are then imposed in regions that are occupied by rigid particles. The resulting solution gives the Brownian motion of the particles. This approach is shown to be viable for the simulation of long time scale diffusive behavior as well as for short time scale dynamics by using two separate solution techniques. Test cases are reported to validate the approach and to establish its efficacy.

---

[*] Corresponding author.

# 1. Motivation and background

The mechanics of intracellular processes is known to play an important role in many biological functions. Biological macromolecules inside the cells are typically sub–micron scale and they are in an aqueous environment. Thermal fluctuations are known to play a crucial role in many functional features of biological processes. The scales of these problems are such that molecular dynamic simulations are typically very expensive. Thus, there has been much interest to develop the so–called "mesoscale" methods that, while coarse–grained, retain the key aspects that are important to simulate Brownian (thermally fluctuating) systems.

Brownian Dynamics (BD) approach, which is based on Langevin equation for particle motion, is one of the most widely used methods to simulate Brownian systems.[1] It uses simplified models for the drag on the particles. Stokesian Dynamics (SD) approach is also based on Langevin equations for Brownian particles, but the hydrodynamic interactions are computed without approximations.[2] However, using these techniques to objects of irregular shapes and to cases where the fluid exhibits varying properties is not straightforward.

Mesoscale methods that fully resolve the hydrodynamic drag and interactions between Brownian particles may be categorized into three types of approaches. First family of approaches, called the Dissipative Particle Dynamics (DPD) methods are primarily based on coarse graining molecular dynamic equations.[3,4] The second approach is based on Lattice Boltzmann (LB) equations.[5,6] The third family of approaches use continuum equations based on Navier-Stokes based fluctuating hydrodynamic (NS-FHD) equations.[7]

NS-FHD equations are the usual Navier–Stokes equations with additional random stress terms that satisfy the Fluctuation–Dissipation Theorem (FDT). The random stress terms model the thermal fluctuations in the fluid. NS-FHD based methods have been applied to solve fluid equations.[8,9] It was shown theoretically that the motion of an isolated Brownian particle computed using fluctuating hydrodynamic equations is consistent with the traditional Langevin description in the long time (dissipative) limit.[10,11] We developed a numerical approach wherein we solved the motion of Brownian particles coupled with NS-FHD equations.[12-14] Other NS-FHD based simulation methods of Brownian particles have also been reported.[15-17]

In our approach the entire fluid–particle domain is considered to be a fluid governed by NS-FHD equations. It is ensured that the 'fluid' occupying the particle domain moves rigidly by adding a rigidity constraint.[14,18-20] A finite volume approach is used. Solution of the coupled system of equations results in the Brownian motion of the particles.[12,14] The key advantages of such an approach is that it can be easily incorporated in existing computational fluid dynamics (CFD) solvers. Problems with irregular particles can be set up without any fundamental difficulty. The translational and rotational Brownian motions are obtained simultaneously. Both long time scale diffusive behavior and short time scale motion can be solved – a result that will be shown in this work. In fact, fluctuating hydrodynamics approach captures the $t^{-3/2}$ algebraic tail in the translational velocity autocorrelation and the $t^{-5/2}$ algebraic tail in the rotational velocity autocorrelation of a sphere. This is consistent with molecular dynamic results.[10] The Langevin equation for particles, used in Brownian dynamics, gives an exponential tail in the velocity autocorrelation function. Thus, fluctuating hydrodynamics based methods are phenomenologically better in comparison for Brownian systems. Our past work focused on dilute systems, spherical particles, and long time diffusive behavior.[14]

Dunweg and Ladd[21] have raised the following relevant issues regarding the use of *finite-difference* methods to simulate systems with thermal fluctuations:

**1)** It is difficult to include thermal fluctuations in the fluid equations in its discretized form.

**2)** Detailed benchmark comparisons with LB approach[5,6] are not available.

**3)** The discretized equations must make sure that mass and momentum are conserved in the domain to machine accuracy, similar to LB methods.

**4)** It is proposed that fluctuating hydrodynamics based methods may not use the incompressible limit because it may not be computationally efficient to work in that limit when there are fluctuations. Incompressibility constraint introduces non–local constraints and requires a Poisson solver which is computationally expensive. Unlike this approach, LB methods use explicit schemes with compressible fluid that are easy to parallelize and are computationally efficient.

The goal of this work is to extend the capability of our approach, to present extensive benchmark results, and to consequently address issues, such as those raised above. Below we summarize the key contributions of this paper vis–a–vis the four issues listed above. The discussion below also serves to provide an overview of work to be presented here, to differentiate this work from our prior work, and to put this work in the wider context of the discussion in the field.

**1)** Unlike previous work,[14] we consider particles at high concentrations. More importantly, results for the rotational diffusion of the particles will be presented. It is noted here that two aspects are important for these results. First, is the assumption of the *symmetric* random stress imposed on staggered CV faces, and second, is presence of random stresses in the *entire* domain, i.e., including the particle region. This ensures that there is no net torque due to the random stress on the main CV and consequently no net torque on the entire computational domain (see sections 2.2 and 3.2).

This approach is general and easy to implement. Thus, it is not difficult to include thermal fluctuations in the fluids equations in its discretized form. This scheme has good spectral behavior (see section 3.1).

**2)** We present new results for benchmark cases compared to our prior work.[14] These include long time scale diffusion results for spherical and ellipsoidal particles (see section 3.2). Non–dilute suspensions with concentrations up to close–packing are reported. Additionally, a time dependent solution scheme is used (see section 4; this temporal scheme was not reported in our prior work) to report short time scale behavior including algebraic tails for translational and rotational autocorrelations. The repertoire of test cases represents extensive validation of our scheme. A simultaneous comparison between NS-FHD and LB based techniques in terms of computational resources is not within the scope of this work. However, we do agree that this task should be undertaken in the future by practitioners in both communities (also see pt. 4 below for additional remarks related to computational efficiency).

**3)** NS-FHD based techniques that use the control volume method, like the one we have developed, are conservative schemes. As such they conserve mass and momentum to machine accuracy.

**4)** Two types of problems are of interest in Brownian systems. First, is to simulate diffusive behavior of the particles on long time scales. Most Brownian or Stokesian dynamic simulations fall in this category. Second, is to simulate short time scale (for example viscous time scale; time scales are explained in section 2.3) behavior where temporal acceleration (inertia) is not negligible.

It is discussed in section 2.3 that the first problem pertaining to long time scale diffusion can be represented by Stokes equations. Solution to Stokes equations (with fluctuating stresses) leads to one instance of Brownian displacement (i.e., diffusive displacement on long time scales). This is an advantage since long time scale behavior can be directly extracted without having to resolve the small time scale behavior. This temporal decoupling is crucial to enhance the efficiency of mesoscale simulation tools compared to say molecular dynamic approach where simulations proceed with time steps on order of short time scales. This approach will be validated by considering variety of test cases in section 3. In this long time limit, the short time scale dynamics pertaining to compressibility effects such as wave motion are not resolved but the incompressibility assumption is reasonable. This is no different from the incompressibility assumption that is inherent in prior techniques such as Stokesian dynamics. Since, this approach requires solution to Stokes equations, temporal schemes such as implicit or explicit approaches are not relevant – there is no time derivative term in the governing equations. As such, there is no additional advantage in using a compressible fluid based approach instead of an incompressible solver – both require Poisson solvers (Note: A Poisson solver for pressure can be avoided by using compressible fluid only if an explicit scheme is used in the presence of time derivative terms in the governing equations). The cost of having to solve Poisson equation is compensated by the ability to directly obtain long time scale solution compared to methods where short time scale dynamics are resolved.

The second type of problem pertains to simulating short time scale (on the order of viscous or wave time scales) dynamics. Here it is essential to keep the time derivative inertia term in the governing equations (see section 2.3). In this case the velocity autocorrelation is affected at small times by using the incompressibility constraint. This has been shown theoretically.[10] However, it is noted in section 4 that algebraic tails in autocorrelation are still recovered, as predicted theoretically, for both translational and rotational motions. Thus, to show algebraic tails in autocorrelation and to establish the efficacy of our constraint–based formulation, using an incompressibility solver is sufficient (which we have used also because we have such a solver in our repertoire). Development of an entirely new compressible solver is not within the scope of this work. However, that is not to be considered a limitation of NS-FHD based schemes. Explicit solvers for compressible fluids with particles and NS-FHD–based equations are being developed by using the MacCormack scheme.[16,22] Such schemes are conceptually equivalent to explicit LB based methods and therefore are expected to be equally efficient for short time scale computations. Our constraint based approach for rigid particle motion can potentially be implemented in the MacCormack schemes for compressible fluids as well. We have not focused on that case here because mesoscale simulations are typically desirable to simulate long time scale diffusive motion of objects that are order ten nanometers to sub–micron scale. It can be verified (see section 2.3) that, on the short viscous or wave motion time scales, nanometer scale or larger objects have negligible motion on the scale of its own size. Thus, short time scale simulations may not be preferable, based on computational cost, in comparison to the long time scale approach based on Stokes equations (section 3).

In this paper the long time dissipative limit is considered first (section 3), which is equivalent to neglecting the inertia terms in the governing equations. Appropriate test cases are presented for the validation. Unsteady simulations are considered in section 4 to test the short time behavior of the velocity autocorrelation function. Conclusions are presented in section 5.

## 2. Navier-Stokes based fluctuating hydrodynamic equations

The governing equations for a fluctuating fluid and their discretization are discussed first; after which the fluid–particle problem is considered.

### 2.1 Differential equations

Let $\Omega$ be the computational domain. Assume that the computational domain is periodic in all directions. The formulation can be extended to non–periodic domains. A Newtonian fluid with constant density and viscosity will be considered for simplicity. The temperature is assumed to be uniform. The fluctuating hydrodynamic equations are

$$\frac{\partial(\rho \mathbf{u})}{\partial t} + \nabla \cdot (\rho \mathbf{u}\mathbf{u}) = -\nabla p + \mu \nabla^2 \mathbf{u} + \nabla \cdot \widetilde{\mathbf{S}} \quad \text{in } \Omega, \tag{1}$$

$$\nabla \cdot \mathbf{u} = 0 \quad \text{in } \Omega, \tag{2}$$

where $\rho$ is the fluid density, $\mu$ is the fluid viscosity, $\mathbf{u}$ is the fluid velocity, $p$ is the dynamic pressure, and $\widetilde{\mathbf{S}}$ is the random stress tensor.[7] $\widetilde{\mathbf{S}}$ is included in the Navier–Stokes equations to model the fluctuations in the fluid at mesoscopic scales (typically from tens of nanometers to microns). $\tilde{\mathbf{S}}$ has following properties[7]

$$\left.\begin{array}{l}\left\langle \tilde{S}_{ij} \right\rangle = 0, \\ \left\langle \tilde{S}_{ik}(x_1,t_1)\tilde{S}_{lm}(x_2,t_2) \right\rangle = 2k_B T \mu \left( \delta_{il}\delta_{km} + \delta_{im}\delta_{kl} \right) \delta(x_1 - x_2)\delta(t_1 - t_2).\end{array}\right\} \quad (3)$$

where $\langle \ \rangle$ denotes averaging over an ensemble, $k_B$ is the Boltzmann constant, $T$ is the temperature of the fluid, and indicial notation has been used. The above equations are in accordance with the fluctuation dissipation theorem for an incompressible fluid.[7,23] The incompressibility assumption is inherent in the earlier work based on BD or SD type simulations. As discussed in section 1 the same assumptions are made here. For a discussion of the fluctuation equations in the presence of constraints, the reader is referred to prior literature.[24,25] The energy equation is not considered because the temperature is assumed to be uniform.

Öttinger & Grmela[23] have shown that the above equations, in the differential form, are thermodynamically consistent. They show this by formulating the governing equations in the GENERIC (General Equation for Non–Equilibrium Reversible/Irreversible Coupling) form. The extension of the equations to varying viscosity and temperature, compressible or non–Newtonian fluids, is also possible.[23] Governing equations (1)–(3) for fluctuating hydrodynamics assume that there is no body torque acting at a point in the fluid. As such, the viscous stress tensor is symmetric. Consequently, the FDT implies that the random stress tensor $\tilde{\mathbf{S}}$ is symmetric.[23]

If $\tilde{\mathbf{S}}$ has, both, a symmetric and an asymmetric part, then it would result in a fluctuating body torque acting at a point in the fluid. In this case the viscous stress tensor will not be symmetric. This problem is not considered in this work.

## 2.2 Spatial discretization

Governing equations (1)–(3) are stochastic. Spatial and temporal discretization of the governing equations should be done such that the discretized equations satisfy the corresponding FDT. It must be noted that even if the differential equations satisfy the FDT it does not imply that the corresponding discretized equations based on central differencing will necessarily satisfy the FDT for the discrete equations.

Thermodynamic consistency of the discrete equations can be ensured if the equations are discretized such that they are in the GENERIC form.[23,26] A systematic derivation of two–dimensional discrete equations that are in the GENERIC form has been presented for the case of a fluid.[9,27] A finite volume Lagrangian discretization based on Voronoi tessellation was used. Prior work[9,27] implies that simple central differencing does not ensure thermodynamically consistent discretized equations. This can be corrected by adding certain terms to the discrete equations. It was also argued that these additional terms may be neglected under certain condition. This will be discussed further when the discretization of equations (1)–(3) is presented below.

A control (finite) volume discretization based on cubic cells is used. A staggered control volume scheme is preferred; the reason for which will be discussed later. A uniform grid, with discretizations $\Delta x = \Delta y = \Delta z = \Delta h$ in the $x$, $y$ and $z$ directions, respectively, is used. The scheme can be extended to non–uniform grids.

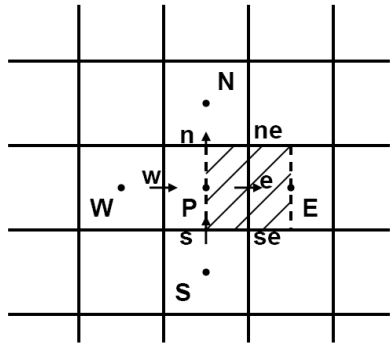

Figure 1. The main control volumes are enclosed by bold lines. A staggered control volume for $x$–velocity is shown by a marked box.

The control volumes are depicted in figure (1). A node P at the center of the main control volume (CV) is surrounded by six neighbors E, W, N, S, T and B. T and B, not shown in figure (1), are the nodes in the $z$–direction above and below the node P, respectively. The six faces of the main CV are named e, w, n, s, t and b. The faces t and b are in the $z$–direction.

The pressure is defined at the center of the main CV's. The $x$–velocities are defined on the e and w faces, $y$–velocities on the n and s faces and $z$–velocities on the t and b faces of the main CV. They are indicated by corresponding arrows in figure (1). The discrete form of the continuity equation is derived by integrating equation (2) over the main control volume.

The *x*–momentum equation is derived by integrating the *x*–component of equation (1) over the staggered CV for *x*–velocity (marked in figure 1). The staggered CV for *x*–velocity has six faces – E, P, ne, se, te and se. The nomenclature is the following – ne indicates the face to the north (n) of location e and so on. The *x*–velocity $u_e$ at the center of the staggered CV is surrounded by six neighboring *x*–velocities. Similarly, the *y*– and *z*–velocity CVs are used to obtain the discrete form of the corresponding momentum equations.

The spatially discretized forms of the stochastic governing equations are obtained by following the derivation of Serrano et al.[27] The continuity equation (2), for a typical main CV, becomes

$$(u^e - u^w) + (v^n - v^s) + (w^t - w^b) = 0, \qquad (4)$$

where *u, v* and *w* denote the *x, y*, and *z* components of velocity **u**, respectively. Superscripts denote the main CV faces on which the velocities are defined. The *x, y* and *z* momentum equations, at typical staggered CV's, are

$$\begin{aligned}d(\rho u^e)\Delta h^3 &= -\sum_{CV\ faces}\left((\rho \mathbf{u}\cdot \mathbf{n}\Delta h^2)u\right)^{CV\ faces} dt + \left\{\sum_{nb}\mu_{eff}\Delta h u^{nb} - 6\mu_{eff}\Delta h u^e\right\} dt \\ &+ (p^P - p^E)\Delta h^2 dt + \sqrt{2k_B T \mu \Delta h}\left\{(d\widetilde{W}_{xx}^E - d\widetilde{W}_{xx}^P) + (d\widetilde{W}_{xy}^{ne} - d\widetilde{W}_{xy}^{se}) + (d\widetilde{W}_{xz}^{te} - d\widetilde{W}_{xz}^{be})\right\},\end{aligned} \qquad (5a)$$

$$\begin{aligned}d(\rho v^n)\Delta h^3 &= -\sum_{CV\ faces}\left((\rho \mathbf{u}\cdot \mathbf{n}\Delta h^2)v\right)^{CV\ faces} dt + \left\{\sum_{nb}\mu_{eff}\Delta h v^{nb} - 6\mu_{eff}\Delta h v^n\right\} dt \\ &+ (p^P - p^N)\Delta h^2 dt + \sqrt{2k_B T \mu \Delta h}\left\{(d\widetilde{W}_{yx}^{en} - d\widetilde{W}_{yx}^{wn}) + (d\widetilde{W}_{yy}^N - d\widetilde{W}_{yy}^P) + (d\widetilde{W}_{yz}^{tn} - d\widetilde{W}_{yz}^{bn})\right\},\end{aligned} \qquad (5b)$$

$$\begin{aligned}d(\rho w^t)\Delta h^3 &= -\sum_{CV\ faces}\left((\rho \mathbf{u}\cdot \mathbf{n}\Delta h^2)w\right)^{CV\ faces} dt + \left\{\sum_{nb}\mu_{eff}\Delta h w^{nb} - 6\mu_{eff}\Delta h w^t\right\} dt \\ &+ (p^P - p^T)\Delta h^2 dt + \sqrt{2k_B T \mu \Delta h}\left\{(d\widetilde{W}_{zx}^{et} - d\widetilde{W}_{zx}^{wt}) + (d\widetilde{W}_{zy}^{nt} - d\widetilde{W}_{zy}^{st}) + (d\widetilde{W}_{zz}^T - d\widetilde{W}_{zz}^P)\right\},\end{aligned} \qquad (5c)$$

where superscript *nb* denotes the six neighboring velocities for the respective staggered CV's. Similarly, superscripts on pressure and $d\widetilde{W}_{ij}$ denote the faces of the respective staggered CV's. *d*( ) denotes differential changes in time *dt*. In equations (5a)–(5c), the first terms on the right hand side are due to convection and the terms after that in brace brackets are due to viscous diffusion.

Equations (5a)–(5c) represent spatially discretized stochastic differential equations (with respect to time) according to the Ito interpretation (see Kloeden & Platen,[28] Chapter 3). $d\widetilde{W}_{ij}$ (= $d\widetilde{S}_{ij} dt$; Ito interpretation implied) are the increments of Wiener processes (see Kloeden & Platen,[28] Chapters 1 & 2) at the respective CV faces. Subscripts on $d\widetilde{W}$ denote the stress components. The properties of $d\widetilde{W}_{ij}$ follow from those of the random stress $\widetilde{\mathbf{S}}$ and are given by

$$\langle d\widetilde{W}_{ij} \rangle = 0, \quad \langle d\widetilde{W}_{ik} d\widetilde{W}_{lm} \rangle = (\delta_{il}\delta_{km} + \delta_{im}\delta_{kl})dt, \tag{6}$$

and $\mu^{eff}$ in equation (5) is[27]

$$\mu_{eff} = \mu\left(1 + \frac{k_B}{2C}\right), \tag{7}$$

where $C$ is the heat capacity of the fluid inside a CV.

In equations (5a)–(5c), the source terms are due to the random stresses (or the Wiener processes). Since we consider a periodic computational domain, it can be easily verified that if we add these source terms over all the control volumes for $x$–velocity then the sum is zero. This implies that there is no net force on the computational domain in the $x$–direction. Similarly, summations in the $y$ and $z$ directions are also zero. This ensures that the mean force on the computational domain due to the random stresses is zero.

It can be verified that the net torque, due to the random stresses, on each *main CV* is zero. This is because the random stress is symmetric. It also follows that net torque on the entire computational domain is zero.

Equations (4)–(7) are in the GENERIC form and obey the FDT.[27] A straightforward central difference discretization of equation (1) will not give the extra term $k_B/2C$ in equation (7). This is the additional term required to satisfy the FDT for the discrete equations. However, this term is typically a small quantity and is proportional to the inverse of the number of molecules of the fluid in the CV.[27] Hence, if it is neglected we get $\mu^{eff} = \mu$. The resulting discrete momentum equations are the same as those obtained by a simple central difference discretization of equation (1). In this case the FDT is not strictly satisfied but the error is small if $k_B/2C$ is small. In this work we will put $\mu^{eff} = \mu$.

A control volume is interpreted as a fluid mesoparticle whose motion is, in general, tracked by an equation of the type $\left(d\mathbf{X}_{CV}/dt\right) = \mathbf{u}$, where $\mathbf{X}_{CV}$ is the node associated with the CV and $\mathbf{u}$ is its velocity.[9,27] Similarly, the deformation can be tracked by considering the motion of each point on the surface of the CV. This is the Lagrangian tracking of the CV's. When Lagrangian tracking is necessary, the uniform grid as shown in figure (1) can be used only at the initial instant. At all subsequent times the grids should be updated to the new location and shape; for example, Serrano & Español[9] and Serrano et al.[27] constructed CV's based on Voronoi tessellation at each time step. This makes the problem computationally intensive. However, it will be noted below that the Lagrangian tracking of CV's can be ignored for the problems of interest in this work.

*2.3 Time discrete approximation*

In this work two types of problems will be considered. In section 3, the long time dissipative limit is considered, which is equivalent to neglecting the time derivative terms (left hand side of equation 5) and the convection terms (first term on the right hand side of equation 5). This results in Stokes' problem. Unsteady simulations will be considered in section 4. The time discrete approximation and the scaling of the governing equations will be considered next.

Problems driven *only* by the random stresses will be considered. A single fluid will be considered at this stage. The numerical solution of fluctuating hydrodynamic equations for fluids will be referred to as Computational Fluctuating Hydrodynamics (CFHD). The simulations in this work will be performed in a fully periodic domain. As an example and for the purposes of presenting the scaling, an implicit time discrete approximation of the governing equations (4)–(6) is given below. The variables are non–dimensionalized by using the following scales

$$\left. \begin{array}{l} L_s \rightarrow \Delta h, \; S \rightarrow \sqrt{2k_B T \mu / \Delta h^3 \Delta t}, \\ V \rightarrow SL_s/\mu = \sqrt{2k_B T / \rho \Delta h^3} \sqrt{T_\mu / \Delta t}, \\ P \rightarrow \mu V / L_s, \end{array} \right\} \quad (8)$$

where $\Delta t$ is the time step and $T_\mu$ is the viscous time scale given by $T_\mu = \rho \Delta h^2 / \mu$. $L_s$ is the length scale based on the grid size which determines the size of the fluid mesoparticle, $S$ is the scale for the random stress, $V$ is the velocity scale, and $P$ is the pressure scale. The scale for the random stress follows from its variance. The velocity scale follows by equating the scales of the viscous and random stresses. The pressure is scaled by the viscous stress. It is noted that since the velocity and pressure are random variables, the square of their scales give the scale of their variance. Thus, the scales are relevant in the context of the variance of the variables. The variance can be computed based on simulations of many ensembles.

The non–dimensionalized governing equations are

$$\left.\begin{array}{l} \left(\dfrac{T_\mu}{\Delta t}\right)\left(\mathbf{u}^{n+1} - \mathbf{u}^n\right) + \left(\dfrac{\rho V L_s}{\mu}\right)\nabla \cdot \left(\mathbf{u}^{n+1}\mathbf{u}^{n+1}\right) = -\nabla p^{n+1} + \nabla^2 \mathbf{u}^{n+1} + \nabla \cdot \left(\Delta \widetilde{\mathbf{W}}\right), \\ \nabla \cdot \mathbf{u}^{n+1} = 0, \\ \left\langle \Delta \widetilde{W}_{ij} \right\rangle = 0, \quad \left\langle \Delta \widetilde{W}_{ik} \Delta \widetilde{W}_{lm} \right\rangle = \left(\delta_{il}\delta_{km} + \delta_{im}\delta_{kl}\right), \end{array}\right\} \quad (9)$$

where same symbols have been retained for the non–dimensional variables. Equation (9) represents a non–dimensionalized time discrete approximation of equations (4)–(6), with $\mu^{eff} = \mu$. As such, the differential symbols for spatial derivatives have been used simply for the convenience of presentation. They should be considered to imply the corresponding central difference discrete operators. $\Delta \widetilde{\mathbf{W}}$ follows from $d\widetilde{\mathbf{W}}$ in equation (5) and is the increment of the Wiener process over time step $\Delta t$. In accordance with the scale of random stresses, $\Delta \widetilde{\mathbf{W}}$ is non–dimensionalized by $\sqrt{\Delta t}$. The same symbol is retained in equation (9) for $\Delta \widetilde{\mathbf{W}}$.

Superscripts in equation (9) denote the time at which the variable is computed. It represents an implicit time discrete approximation of equations (4)–(6). Based on the Ito–Taylor expansion, this is called the implicit strong Taylor scheme of order ½.[28] The convergence of time discretized stochastic differential equations is different from that of the deterministic differential equations. Two types of convergence, namely strong convergence and weak convergence, are relevant in case of stochastic equations. The implicit scheme above has a strong convergence of order ½ and a weak convergence of order 1 with respect to the time step $\Delta t$.[28] This is briefly explained below in the interest of clarity and completeness of the presentation.

Consider an exact solution, with respect to time, of the spatially discretized equations (4)–(6). Let the mesh size $\Delta h$ be fixed. Let $\mathbf{u}(\mathbf{X}_A, t_T)$ be the exact solution at location $\mathbf{X}_A$ at time $t_T$. Let $\mathbf{u}_N^{\Delta t}(\mathbf{X}_A, t_T)$ be the solution of the time discretized equation, corresponding to a time step $\Delta t$, at the same location $\mathbf{X}_A$ and time $t_T$. In the current discussion $\mathbf{u}$ and $\mathbf{u}_N^{\Delta t}$ are considered to be dimensional variables. Let the initial conditions as well as the Wiener processes $\widetilde{W}_{ij}$, in computing the exact and numerical solutions, be the same. Calculate the exact and the numerical solutions (for a given $\Delta t$) for many trajectories (or ensembles) by using different sets of the Weiner processes but the same initial condition. Compute the error, $\langle |\mathbf{u}(\mathbf{X}_A, t_T) - \mathbf{u}_N^{\Delta t}(\mathbf{X}_A, t_T)| \rangle$, based on the ensemble average. Compute similar errors for different values of $\Delta t$. A strong convergence of order ½ implies that[28]

$$\langle |\mathbf{u}(\mathbf{X}_A, t_T) - \mathbf{u}_N^{\Delta t}(\mathbf{X}_A, t_T)| \rangle \sim K \Delta t^{1/2}, \tag{10}$$

where $K$ is some constant and $|\ |$ implies absolute value. A weak convergence of order 1 implies that[28]

$$|\langle \mathbf{u}(\mathbf{X}_A, t_T) \rangle - \langle \mathbf{u}_N^{\Delta t}(\mathbf{X}_A, t_T) \rangle| \sim K \Delta t. \tag{11}$$

In section (4), where unsteady simulations with particles are considered, an implicit fractional time stepping scheme is used.

The scaling in the Stokes and unsteady simulations will be discussed below based on equation (9).

***Unsteady simulations:*** In case of unsteady simulations, the time step $\Delta t$ is *at least* of the order of the viscous times scale $T_\mu$, i.e., $\Delta t \sim T_\mu$. Thus, the time derivative term is at least of the same order as the viscous and random stress terms. Equation (8) implies that the velocity scale $V \sim \sqrt{2k_B T / \rho \Delta h^3}$. Since velocity is a random variable, as noted above, it follows that $\left\langle \rho \Delta h^3 V^2 / 2 \right\rangle \sim k_B T$, i.e., the mean kinetic energy of the fluid mesoparticle (or the CV) is of the order of $k_B T$. This is in agreement with the equipartition theorem for a system at equilibrium.

The Reynolds number, $\rho V L_s / \mu$, gives the scale of the convection term. Using the expression for the velocity scale we get $\rho V L_s / \mu \sim \sqrt{2\rho k_B T / \mu^2 \Delta h}$. The convection term is negligible compared to the remaining terms if

$$\sqrt{\frac{2\rho k_B T}{\mu^2 \Delta h}} \ll 1 \text{ i.e. if } \Delta h \gg \frac{2\rho k_B T}{\mu^2}. \tag{12}$$

For water at 300K, equation (12) implies $\Delta h \gg 8.28 \times 10^{-12}$ m. The intermolecular distance for water is of 3–4A$^\circ$. The fluctuating hydrodynamic equations are relevant for scales where the continuum equations are reasonable, i.e., $\Delta h$ should be greater than 5nm. Thus, equation (12) is easily satisfied for cases of relevance. The convection term is therefore negligible in equation (9). Similar conclusion is valid for typical problems in the context of Brownian motion. In this work the convection term will be ignored. This conclusion is also in agreement with the remarks of Hauge & Martin–Löf[10] (see page 263 of their paper).

For $\Delta t \sim T_\mu$, the scale of the displacement of the fluid mesoparticles is $\Delta \mathbf{X}_{CV} \sim V \Delta t = \sqrt{2k_B T / \rho \Delta h^3} \Delta t$, where $\Delta \mathbf{X}_{CV}$ is dimensional. Thus, the mean square displacement scales as $\left\langle \Delta \mathbf{X}_{CV}^2 \right\rangle \sim \left( 2k_B T / \rho \Delta h^3 \right) \Delta t^2$. This is the scaling for the short time Brownian displacement[29] of the fluid mesoparticles. The Brownian displacement relative to the length scale $\Delta h$ is $\Delta \mathbf{X}_{CV} / \Delta h \sim \rho V L_s / \mu \ll 1$ (ref. discussion following equation 12). Physically, this implies that the displacement of the fluid mesoparticle in time $\Delta t$ (which is of the order of the viscous time scale $T_\mu$) is negligible compared to its size $\Delta h$. Therefore, in this work the Lagrangian tracking of the fluid mesoparticles is not done.

The unsteady simulations are considered in section 4.

***Stokes simulations:*** These simulations are relevant when $\Delta t \gg T_\mu$. In this case, the viscous effects dominate and the inertia terms (the entire left hand side of the momentum equation in equation 9) are neglected. The resulting Stokes' problem can be solved for variable **u**. However, **u** is not the actual velocity of the fluid at any instant. It should be interpreted as the apparent velocity of the fluid mesoparticle based on its Brownian displacement, $\mathbf{X}_{CV}(\Delta t) - \mathbf{X}_{CV}(0)$, in time $\Delta t$, i.e., $\mathbf{u} = (\mathbf{X}_{CV}(\Delta t) - \mathbf{X}_{CV}(0))/\Delta t$.[14] This equation is identical to the discretization of $\left( d\mathbf{X}_{CV}/dt \right) = \mathbf{u}$ for the Lagrangian tracking of the CV's. However, in Stokes' problem the time interval $\Delta t$ is sufficiently large so that **u** is not an approximation of the actual velocity of the fluid mesoparticles at intermediate time instants. Equation (8) implies that, in Stokes' problem, the velocity scale $V \sim \sqrt{2k_BT/\rho \Delta h^3}\sqrt{T_\mu/\Delta t} \ll \sqrt{2k_BT/\rho \Delta h^3}$, where $\sqrt{2k_BT/\rho \Delta h^3}$ is the instantaneous velocity scale according to the equipartition of energy. Thus, the scale of the apparent velocity is less than the scale of the actual velocity.

The scale of the displacement of the fluid mesoparticles is $\Delta \mathbf{X}_{CV} \sim V\Delta t = \sqrt{2k_BT/\rho \Delta h^3}\sqrt{T_\mu/\Delta t}\Delta t$, where $\Delta \mathbf{X}_{CV}$ is dimensional. Thus, the mean square displacement, in this case, scales as $\langle \Delta \mathbf{X}_{CV}^2 \rangle \sim \left( 2k_BT/\mu \Delta h \right)\Delta t$. Once again, this is the expected long time Brownian displacement of the fluid mesoparticles.[29] The Brownian displacement relative to the length scale $\Delta h$ is $\Delta \mathbf{X}_{CV}/\Delta h \sim \sqrt{2k_BT\Delta t/\mu \Delta h^3}$. If the Brownian displacement is to be of the order of $\Delta h$, then this implies that $\Delta t \sim \mu \Delta h^3/2k_BT$, is required. For the Stokes simulations considered in this work, the Lagrangian tracking of the fluid mesoparticles is not relevant. This will become clear when Stokes' problems are considered in the next section.

So far the discussion has focused on the governing equations of the fluid. The equations for the fluid–particle problem are discussed in sections 3.2 and 4.

## 3. Stokes' problem with fluctuations

In section 3.1, the approach is tested for a single fluid first and then the fluid–particle problem is considered in section 3.2.

### *3.1 Computational fluctuating hydrodynamics (CFHD): The Stokes limit*

A single fluid is considered in a periodic domain of size $L \times L \times L$. Governing equations for Stokes' problem are given by equations (1)–(3), except that the inertia terms in equation (1) are neglected and $\delta(t_1 - t_2)$ is replaced by $1/\Delta t$ in equation (3) to represent the time discrete approximation as discussed in section 2.3. The fluid is driven by the random stresses and **u** should be interpreted as the apparent velocity as discussed in section 2.3. An analytical solution to the problem, based on Fourier transforms, will be briefly presented; then the numerical solution will be compared to it.

Consider a vector **f** that is a function of **x**. The Fourier transform and the inverse Fourier transform are defined as

$$\mathbf{f}^{\mathbf{k}} = \frac{1}{L^3} \int_0^L \int_0^L \int_0^L \mathbf{f}(\mathbf{x}) e^{-2\pi i \mathbf{k} \cdot \mathbf{x}/L} d\mathbf{x}, \quad \text{and} \quad \mathbf{f}(\mathbf{x}) = \sum_{\mathbf{k}} \mathbf{f}^{\mathbf{k}} e^{2\pi i \mathbf{k} \cdot \mathbf{x}/L}, \tag{13}$$

respectively, where **k** is the vector of wavenumbers. Each component of **k** takes positive and negative values. The magnitudes range from 0 to ∞. The governing equations (1)–(3), written in terms of the Fourier modes are

$$\left. \begin{aligned} & -i\mathbf{k} p^{\mathbf{k}} - \mu |\mathbf{k}|^2 \mathbf{u}^{\mathbf{k}} \frac{2\pi}{L} + i\mathbf{k} \cdot \widetilde{\mathbf{S}}^{\mathbf{k}} = \mathbf{0}, \\ & \mathbf{k} \cdot \mathbf{u}^{\mathbf{k}} = 0, \\ & \left\langle \widetilde{S}_{ij}^{\mathbf{k}} \right\rangle = 0, \quad \left\langle \widetilde{S}_{ik}^{\mathbf{k}} \widetilde{S}_{lm}^{\mathbf{k}'} \right\rangle = \begin{cases} 0, & \mathbf{k} \neq \mathbf{k}', \\ \dfrac{2k_B T \mu}{L^3 \Delta t} (\delta_{il}\delta_{km} + \delta_{im}\delta_{kl}), & \mathbf{k} = -\mathbf{k}'. \end{cases} \end{aligned} \right\} \tag{14}$$

The solution for $\mathbf{u}^{\mathbf{k}}$ is obtained by solving equation (14):

$$\langle u_i^{\mathbf{k}} \rangle = 0, \quad \langle u_i^{\mathbf{k}} u_m^{\mathbf{k}'} \rangle = \begin{cases} 0, & \mathbf{k} \ne \mathbf{k}', \\ \dfrac{k_B T}{2\pi^2 \mu L \Delta t} \left( \dfrac{\delta_{im}}{|\mathbf{k}|^2} - \dfrac{k_i k_m}{|\mathbf{k}|^4} \right), & \mathbf{k} = -\mathbf{k}'. \end{cases} \qquad (15)$$

Equation (15) can be used to verify the numerical results.

The same problem, with a periodic computational domain of size $L \times L \times L$, was considered for numerical simulations. The governing equations are given by equation (9), except that the inertia terms are neglected in the momentum equation. The superscripts $n+1$ are not relevant in this problem and $\mathbf{u}$ is the apparent velocity. The scales for non–dimensionalization are as given in equation (8). A staggered grid was generated with $N = L/\Delta h = 40$ control volumes in each direction. The components of the random stresses at different locations were generated from a Gaussian random number generator with the desired mean and variance. The resulting Stokes' problem was solved by the SIMPLER algorithm.[30] A Fast Fourier Transform (FFT) solver (FISHPAK) was used to solve the apparent velocity and pressure equations. It was ensured that the net 'momentum' based on the apparent velocity in the periodic domain was zero. This is because there is no net random 'force' on the computational domain. One simulation was considered as one realization. Several realizations, constituting an ensemble, were solved. For each realization a different initial seed was assigned to the Gaussian random number generator for random stresses. This ensured that each realization was different. A discrete Fourier transform of the apparent velocity field of a given realization was done, after the solution was obtained. This gave $\mathbf{u}^{\mathbf{k}}$ at each realization.

To compare the numerical and analytical results, we considered $\mathbf{k} \equiv (0, k, 0)$ and $\mathbf{k}' = -\mathbf{k}$ in equation (15) for $i = m = 1$ (i.e., the $x$ component of $\mathbf{u}$). Equation (15) implies $\langle u^{(0,k,0)} u^{(0,-k,0)} \rangle = \dfrac{k_B T}{2\pi^2 k^2 \mu L \Delta t}$, where $u$ is the $x$ component of $\mathbf{u}$. Since the numerical results are non–dimensional, we non–dimensionalized this analytical result with the same velocity scale as in equation (8). The analytical result in non–dimensional form is:

$$\langle u^{(0,k,0)} u^{(0,-k,0)} \rangle = \dfrac{1}{4\pi^2 k^2 N}, \qquad (16)$$

where $u$ is non–dimensional and $N = L/\Delta h$ are the number of control volumes in each direction. Equations similar to (16) can also be obtained for the other apparent velocity components.

Two remarks are to be noted. First, in the numerical simulations, the wavenumbers are finite. Thus, the magnitude of each component of $\mathbf{k}$ goes from 0 to $N/2$, where the largest value is determined by the grid resolution. When $\mathbf{k} = \mathbf{0}$ we have $\mathbf{u}^\mathbf{k} = \mathbf{0}$ because the net 'momentum' based on the apparent velocity is zero. Therefore, it is useful to compare equation (16) with the numerical results only for $1 \leq k \leq N/2$.

Second, the waves corresponding to different wavenumbers are not resolved exactly due the discretization process. The error is primarily introduced at larger wavenumbers, i.e., on the scale of the grid size. In this work the derivatives are evaluated at locations that are staggered by a half–cell from the nodes at which the variables are defined. This is useful in reducing the error at larger wavenumbers.[31] Thus, the staggered grid approach was preferred in this work compared to the co–located grid approach. To account for this error the following equation should be used instead of equation (16)

$$\left\langle u^{(0,k,0)} u^{(0,-k,0)} \right\rangle = \frac{1}{4\pi^2 \left(k_{mod}(k)\right)^2 N}, \quad \text{where } k_{mod}(k) = \frac{N}{\pi}\sin\left(\frac{\pi k}{N}\right). \quad (17)$$

$k_{mod}$ is the modified wavenumber corresponding to the wavenumber $k$.[31] The numerical results should be compared to equation (17).

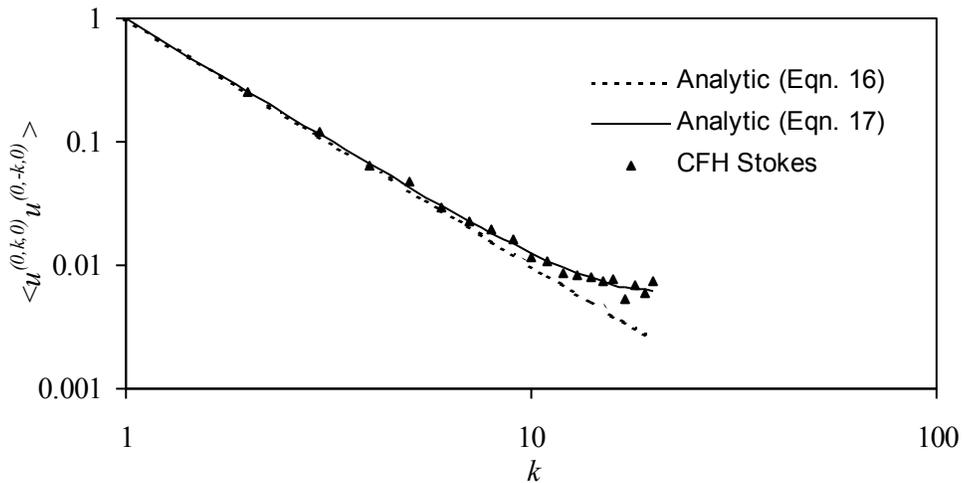

Figure 2. Plot of $\left\langle u^{(0,k,0)} u^{(0,-k,0)} \right\rangle$ vs. $k$ for a single fluid case. $u$ is non–dimensional.

Figure (2) shows a plot of $\langle u^{(0,k,0)} u^{(0,-k,0)} \rangle$ vs. $k$, based on equations (16) and (17). The numerical data are also shown. It is seen that the numerical data are in excellent agreement with the plot due to equation (17). The numerical plot (and equation 17) deviates from the plot based on equation (16) at larger wavenumbers. This deviation would be much more pronounced for a collocated grid based solution of the fluid equations.[31] Ideally, a numerical scheme should be such that the data are as close to the plot due to equation (16) as possible. The results based on spectral methods would be identical to those due to equation (16). However, for fluid–particle problems, to be considered next, it is not straightforward to incorporate the effect of particles within a spectral scheme.

### 3.2 Brownian motion of particles: FIMAT Stokes simulations

Consider a periodic computational domain $\Omega$ in which a particle occupies domain P. One particle is considered here only for the simplicity of presentation. In fact, in section 3.2.3, many particle results are reported. The particle can be of any general shape; in this work we will consider either spherical or ellipsoidal particles. The fluid and the particle densities are assumed to be same for simplicity. To solve this problem the entire fluid–particle domain is assumed to be a fluid. The momentum equation for the fluid–particle Stokes' problem is

$$-\nabla p + \mu \nabla^2 \mathbf{u} + \mathbf{F} + \nabla \cdot \widetilde{\mathbf{S}} = \mathbf{0}. \tag{18}$$

Equation (18) is applicable in the entire domain. It is the same as the equation in Stokes' problem for the single fluid case, except for the extra term $\mathbf{F}$. The properties of $\widetilde{\mathbf{S}}$ are as given in equation (3) except that $\delta(t_1 - t_2)$ is replaced by $1/\Delta t$ to represent the time discrete approximation as discussed in section 2.3. $\widetilde{\mathbf{S}}$ is symmetric and is generated in the entire domain including the particle region. $\mathbf{u}$ is divergence free in the entire domain (equation 2).

$\mathbf{F}$ is non–zero only in the particle domain and is a direct consequence of the rigidity constraint in the particle domain:[18,20]

$$\nabla \cdot \mathbf{D}[\mathbf{u}] = \mathbf{0} \text{ in P, and } \mathbf{D}[\mathbf{u}] \cdot \mathbf{n} = \mathbf{0} \text{ on } \partial P, \tag{19}$$

where $\mathbf{n}$ is a unit outward normal on the particle surface. Equation (19) ensures that the deformation–rate tensor $\mathbf{D}[\mathbf{u}] = \frac{1}{2}(\nabla \mathbf{u} + \nabla \mathbf{u}^T) = \mathbf{0}$ in P. Thus, the 'fluid' in the particle domain is constrained to move rigidly as required. Equation (19) represents three scalar constraint equations at a point in the particle domain. They give rise to a force **F** in the particle domain similar to the presence of pressure due to the incompressibility constraint.[18] This is the Distributed Lagrange Multiplier (DLM) approach for particulate flows.[18,32]

Solution of equations (18), (19), (2), and (3) gives the apparent velocity field **u** in the entire domain. These simulations will be called the Fluctuating Immersed MATerial (FIMAT) dynamics Stokes simulations. Central differencing is used to spatially discretize the equations (section 2.2). The solution procedure is given by Sharma & Patankar.[14,19] The particle translational and angular apparent velocities, **U** and ω, respectively, can then be computed by

$$M\mathbf{U} = \int_P \rho \mathbf{u} d\mathbf{x} \quad \text{and} \quad \mathbf{I}_p \boldsymbol{\omega} = \int_P \mathbf{r} \times \rho \mathbf{u} d\mathbf{x}, \qquad (20)$$

where **r** is the position vector of a point with respect to the centroid of the particle, $\mathbf{I}_p$ is the moment of inertia of the particle and M is the mass. As discussed in section 2.3, **U** and ω should be interpreted as the apparent velocities based on the translational and angular Brownian diffusion of the particles.[14] Results based on FIMAT Stokes simulations for fluid–particle flows will be presented sections 3.2.1–3.2.3.

*3.2.1 Single sphere in a periodic domain*

A single spherical particle of diameter $d$ was placed at the centre of a cubic periodic domain of size $L \times L \times L$. Given $d$ and $L$, the volume fraction $\phi$ of the sphere in the domain is given by $\phi = \pi d^3 / 6L^3$. $N = L/\Delta h = 40$ control volumes were chosen in each direction (this choice was based on grid refinement studies in the earlier work[14]). Random stresses were generated in the entire domain after which the fluid–particle FIMAT Stokes' problem was solved. It was ensured that the net 'momentum' based on the apparent velocity in the entire domain is zero. One simulation was considered as one realization. Ensemble averaging was done based on several realizations. The simulations were performed by solving non–dimensional equations with scales as given in equation (8).

It is also noted here that quasi–steady Stokes simulations (where the particle positions are updated after the solution of the Stokes problem), similar to the BD or SD simulations, can be done using our current method without any fundamental difficulty. However, in this work an ensemble average based approach is preferred for the purpose of validation.

In a given realization, the apparent velocities **U** and ω were computed according to equation (20). Since the apparent velocity should be interpreted in terms of the long time Brownian displacement, the variance of **U** is related to the Brownian diffusion $D_T$, which in turn is related to the translational drag coefficient $K_T$ as follows

$$D_T = \frac{\langle |\mathbf{U}|^2 \rangle \Delta t}{6} = \frac{k_B T}{3\pi \mu d K_T} \Rightarrow K_T = \frac{6 k_B T}{3\pi \mu d \Delta t \langle |\mathbf{U}|^2 \rangle}. \qquad (21)$$

In terms of the non–dimensional variables we get

$$K_T = \frac{1}{\pi d^* \langle |\mathbf{U}^*|^2 \rangle}, \qquad (22)$$

where superscript * implies non–dimensional variables. The scales are chosen as in equation (8). Similarly, the rotational drag coefficient $K_R$ is given by[33]

$$K_R = \frac{6 k_B T}{\pi \mu d^3 \Delta t \langle |\boldsymbol{\omega}|^2 \rangle} = \frac{3}{\pi d^{*3} \langle |\boldsymbol{\omega}^*|^2 \rangle}. \qquad (23)$$

The apparent angular velocity is interpreted in terms of the angular diffusion Δθ, i.e., ω = Δθ/Δt. Equations (22) and (23) were used to compute the drag coefficients from the numerical simulation at a given volume fraction $\phi$.

Figures (3a) and (3b) show histograms of the apparent velocities of the sphere based on 900 realizations at $\phi = 0.008$. It is compared to the analytical Gaussian distribution. The analytical frequency distribution has zero mean in both cases. For the translational case, the variance is $1/\pi d^* K_{T,A}$ (equation 22), where $K_{T,A}$ is the analytical value of the drag coefficient from Hasimoto[34] and Zick & Homsy.[35] For the rotational case the variance is $3/\pi d^{*3} K_{R,A}$ (equation 23), where $K_{R,A}$ is the analytical value by Ladd.[36]

Figure (4) shows the translation drag coefficient $K_T$, calculated according to equation (22), as a function of the volume fraction. The results are compared with the analytical values of Zick & Homsy.[35] The error bars are drawn based on a chi–square distribution for the variance.[14]

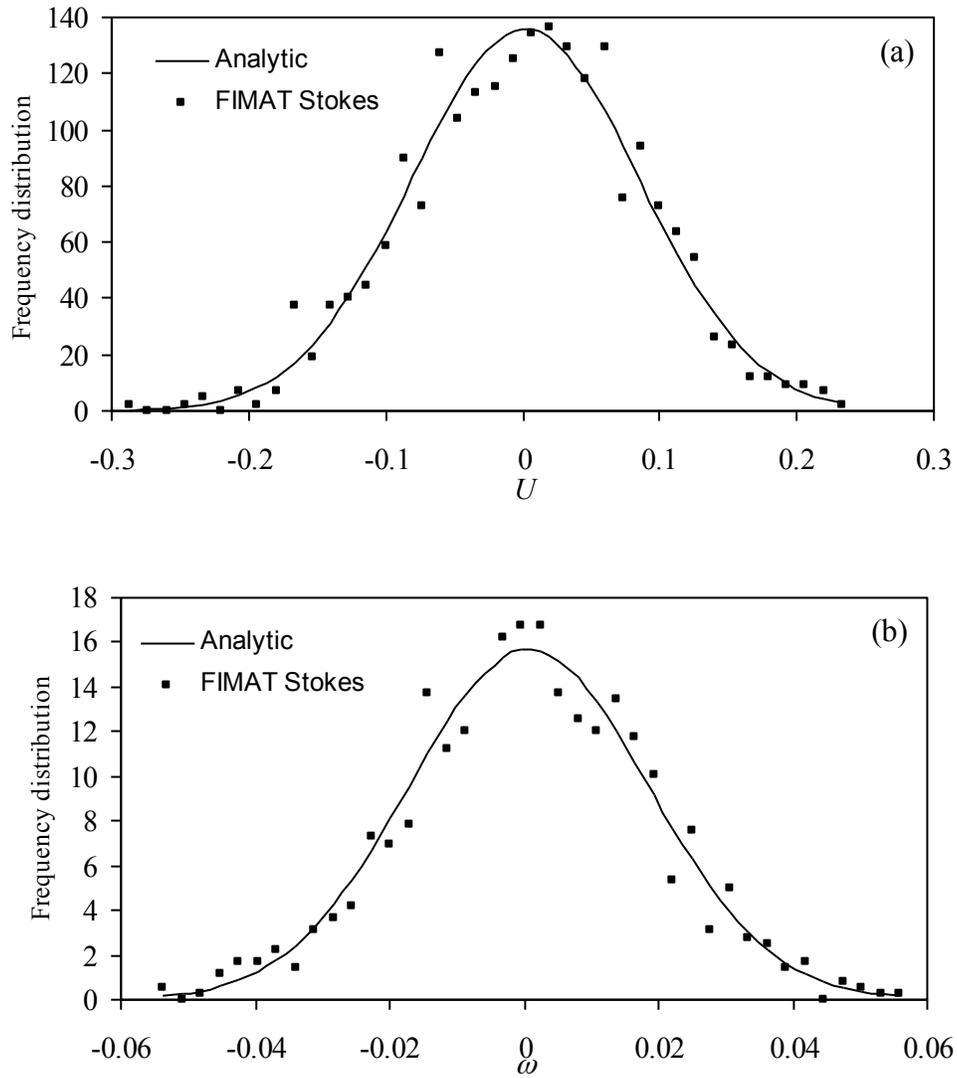

Figure 3. Comparison of the frequency distribution of the apparent velocity (non–dimensional) of the sphere with the analytical Gaussian distribution ($\phi=0.008$): (a) Translational velocity (b) Rotational velocity.

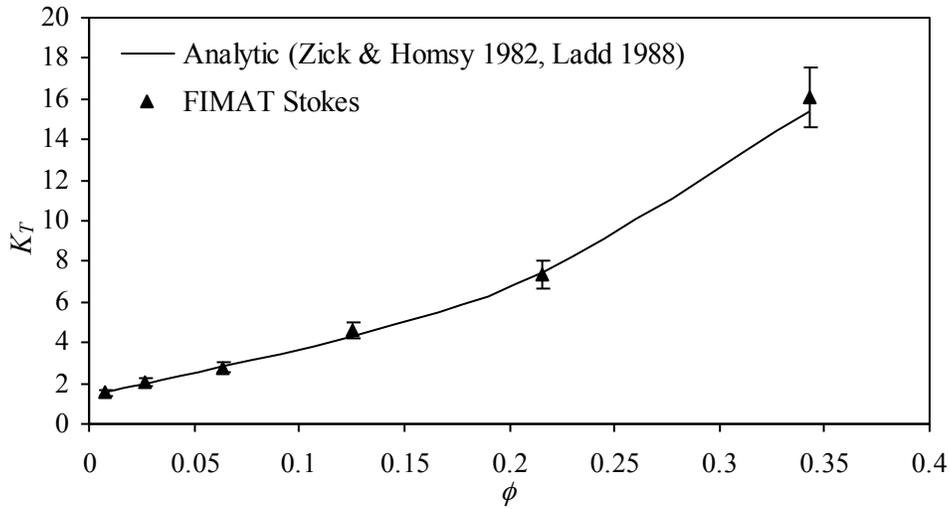

Figure 4. Plot of the translational drag coefficient of a single sphere in a periodic domain as a function of the volume fraction.

Figure (5) shows the results for the rotational drag coefficient $K_R$, calculated using equation (23). These results are compared to the analytical values by Ladd.[37] In all the cases we see that the agreement is good.

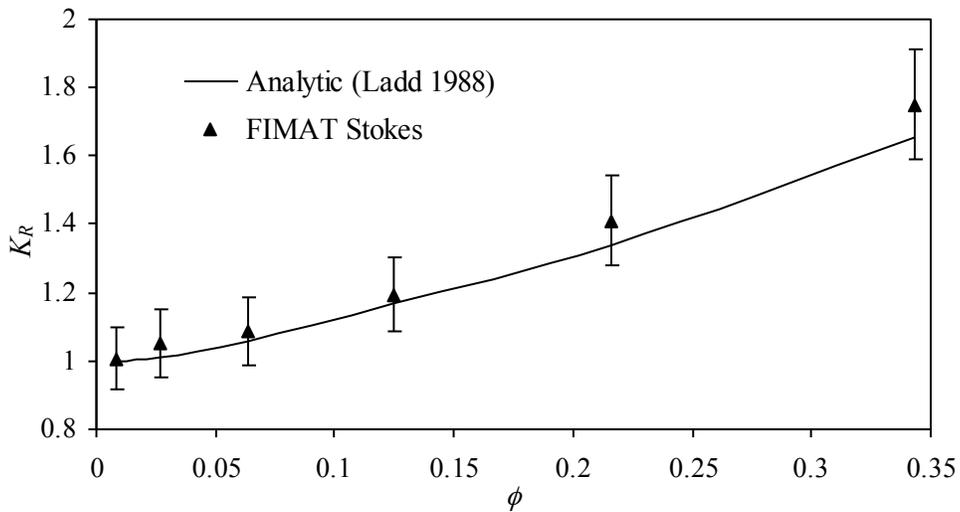

Figure 5. Plot of the rotational drag coefficient of a single sphere in a periodic domain as a function of the volume fraction.

As the close packing limit is approached the lubrication layer between the spheres is not accurately resolved. Due to the presence of the periodic boundary condition, the velocities of the sphere surfaces on either side of the lubrication layer are in the opposite direction. Thus, there can be a sharp gradient of the fluid velocity that is not accurately resolved due to the smearing of the discretized particle boundaries. This problem is common to most of the current fully resolved simulation (FRS) methods for particulate flows.

*3.2.2 Single ellipsoid in a periodic domain*

The problem of ellipsoidal particles is considered next. The translational and rotational drag on an ellipsoidal particle is often quantified in terms of effective radii. Let $F_p$ be the force on an ellipsoid along one of its principle axes and $U$ be its resultant velocity in that direction (assume that there are no fluctuations). Then the effective radius $R_{te}$ for translational motion with respect to the chosen principle direction is defined by[38]

$$R_{te} = \frac{F_p}{6\pi\mu U}. \tag{24}$$

Similarly, the effective radius for the rotational drag is defined by

$$R_{re} = \left(\frac{T_p}{8\pi\mu\omega}\right)^{1/3} \tag{25}$$

where $T_p$ is the torque applied on the ellipsoid with respect to the chosen principle direction and $\omega$ is the corresponding angular velocity (again assume that there are no fluctuations). In general, for an ellipsoid, $R_{te}$ and $R_{re}$ are different in the three principle directions.

The effective radii can also be computed based on the FIMAT Stokes simulations of ellipsoids. For the translational case the effective radius $R_{te}$ is given by

$$D_{te} = \frac{\langle u^2 \rangle \Delta t}{2} = \frac{k_B T}{6\pi\mu R_{te}} \Rightarrow R_{te} = \frac{k_B T}{3\pi\mu \langle u^2 \rangle \Delta t}, \tag{26}$$

where $D_{te}$ is the translational diffusion with respect to one of the principle directions and $u$ is the apparent translational velocity based on the long time Brownian displacement (section 2.3) in that direction. For example, the effective radius with respect to the $x$ direction is obtained if $\langle u^2 \rangle$ is based on the $x$ component of the random apparent velocity of the ellipsoid. Similarly, $R_{re}$ is given by

$$D_{re} = \frac{\langle \omega^2 \rangle \Delta t}{2} = \frac{k_B T}{8\pi\mu R_{re}^3} \Rightarrow R_{re} = \left( \frac{k_B T}{4\pi\mu \langle \omega^2 \rangle \Delta t} \right)^{1/3}, \qquad (27)$$

where $D_{re}$ is the rotational diffusion respect to one of the principle directions and $\omega$ is the corresponding apparent rotational velocity.[33] As discussed in section 3.2.1, the apparent angular velocity is interpreted in terms of the angular diffusion $\Delta\theta$, i.e., $\omega = \Delta\theta/\Delta t$.

$\langle u^2 \rangle$ and $\langle \omega^2 \rangle$ can be computed based on the FIMAT Stokes simulations as discussed in earlier sections. However, to our knowledge, the analytical values of $R_{te}$ and $R_{re}$, for an ellipsoid in a periodic domain, are not available to compare them with the fluctuation results. Lamb[38] has given expressions for $R_{te}$, for an ellipsoidal particle in an infinite domain. These results are not applicable for the periodic domain. Hence, in this work $R_{te}$ and $R_{re}$ were first computed according to equations (24) and (25), where there were no fluctuations in the fluid. These values were then compared with the fluctuation–based calculations.

A single ellipsoidal particle, with semiaxes $a$, $b$ and $c$ in the $x$, $y$ and $z$ directions, respectively, was placed at the center of a cubic periodic domain of size $L \times L \times L$. An oblate ellipsoid was considered, where $a = b = 2c$. $L = 5a$, which implies $\phi = 0.01676$. $N = L/\Delta h = 40$ control volumes were chosen in each direction. The following simulations were performed:

**1.** To calculate $R_{te}$ according to equation (24), Stokes' problem was considered with no fluctuations in the fluid. The flow was driven by a force $F_p$ applied at the centroid of the ellipsoid along one of the principle directions. To ensure that the net force is zero on the periodic computational domain, a constant body force $f_B = F_p/L^3$ per unit volume was applied everywhere in the computational domain in the opposite direction. The resulting Stokes' problem was solved as described by Sharma & Patankar.[19] The ellipsoidal particle velocity was non–zero and equal to $U$ in the direction of $F_p$. Equation (24) gave the value of $R_{te}$ with respect to the chosen principle direction.

**2.** To calculate $R_{re}$, once again Stokes' problem was considered with no fluctuations in the fluid. The flow was driven by a torque $T$, applied on the ellipsoid, with respect to one of the principle directions. This was achieved by imposing equal and opposite forces at two points (also see Sharma & Patankar[19]), on the ellipsoid, located symmetrically with respect to its centroid. The resulting Stokes' problem was solved and the angular velocity $\omega$ with respect to the chosen axis of rotation was calculated. Equation (25) gave the value of $R_{te}$ with respect to the chosen principle direction.

3. Lastly, a FIMAT Stokes' problem was solved similar to the sphere problem in section 3.2.1. The apparent translational and angular velocities at each realization were calculated. Ensemble averaging was done based on several realizations and the values of $R_{te}$ and $R_{re}$ were calculated with respect to all the principle directions.

All simulations were done with non–dimensional variables. Table (1) shows the comparison between the effective radii obtained from the FIMAT Stokes and non–Brownian Stokes simulations for the same geometry and volume fraction (0.01676). The agreement is good. The values with respect to the $x$ and $y$ directions should be same due to the symmetry of the ellipsoid. The results are consistent with this requirement. For an ellipsoid in an infinite domain, $R_{te}/a$ = 0.7925 in the $x$ and $y$ directions and $R_{te}/a$ = 0.9055 in the $z$ direction.[38] As expected the values from the simulations are greater in the periodic domain.

|  | $R_{te}/a$ | $R_{te}/a$ | $R_{re}/a$ | $R_{re}/a$ |
|  | Non–Brownian | FIMAT | Non–Brownian | FIMAT |
| --- | --- | --- | --- | --- |
| $x$–direction | 1.2650 | 1.3410 | 0.8110 | 0.8275 |
| $y$–direction | 1.2650 | 1.3460 | 0.8110 | 0.8020 |
| $z$–direction | 1.6445 | 1.7935 | 0.8790 | 0.9005 |

**Table 1.** Effective radii representing the drag on an ellipsoid from non–Brownian and FIMAT Stokes simulations ($\phi$ = 0.01676).

*3.2.3 Many spheres in a periodic domain*

In this section, many spherical particles are considered in a periodic domain. The diffusivity of these spheres is more than that of a single sphere in a periodic domain (section 3.2.1) at the same volume fraction. This is because the individual spheres have the freedom to move relative to each other. Ladd[36] presented results, based on non–Brownian calculations, for self–diffusion $D_s$ and collective mobility $\mu_c$ of spheres in a periodic domain. In this section, FIMAT Stokes simulations will be done to compare these quantities to those by Ladd.[36]

The self diffusion $D_s$, calculated from FIMAT Stokes simulations, is given by[36]

$$D_s = \frac{\Delta t}{6N_P} \sum_{i=1}^{N_P} \langle |\mathbf{U}_i|^2 \rangle, \qquad (28)$$

where $N_P$ are the number of spheres in the domain. $\mathbf{U}_i$ is the apparent translational velocity of sphere $i$ obtained from FIMAT Stokes simulations. The collective mobility is given by[36]

$$\mu_c = \frac{\Delta t}{6k_B T N_P} \text{tr}\left( \sum_{i,j=1}^{N_P} \langle \mathbf{U}_i \mathbf{U}_j \rangle \right), \qquad (29)$$

where tr denotes a trace of the 3×3 tensor due to $\mathbf{U}_i\mathbf{U}_j$.

First, we considered $N_P = 16$ spheres in a cubic periodic box of side $L$. $N = 40$ was used. All the spheres had the same diameter $d$. Three values of $d/L$ were considered corresponding to volume fractions of $\phi = 0.05$, 0.15 and 0.25. For a given value of $\phi$, the 16 spheres were arranged in a random configuration. Random stresses were generated and the FIMAT Stokes' problem was solved. 450 realizations were simulated, by keeping the random configuration the same, for each volume fraction, respectively. $D_s$ and $\mu_c$ were calculated based on ensemble averaging according to equations (28) and (29).

The computed values are shown in table (2). $D_o = k_B T/3\pi\mu d$ is the diffusion, and $\mu_o = 1/3\pi\mu d$ is the mobility, of a single sphere in an infinite domain. Values based on two different random configurations were computed at $\phi = 0.15$ and 0.25. In table (2), the computed values are compared with those by Ladd.[36] The values by Ladd are based on 16 spheres and are at the same volume fractions as in the simulations. However, his results are based on averaging the values of $D_s$ and $\mu_c$ for 100 different random configurations. In our simulations, we have listed values of $D_s$ and $\mu_c$ for a specific random configuration, to save the computational time. Yet, the agreement is found to be reasonable.

Table (2) also shows the diffusivity of a single sphere in a periodic domain (section 3.2.1) at the same volume fraction as many spheres case. It is seen that in the many spheres case the diffusivity is greater because the spheres can move relative to one another. This is in agreement with the conclusions of Ladd.[36]

| $\phi$ | $D_s/D_o$ Ladd (1990) | $D_s/D_o$ FIMAT | $D_T/D_o$ One sphere | $\mu_c/\mu_o$ Ladd (1990) | $\mu_c/\mu_o$ FIMAT |
|---|---|---|---|---|---|
| 0.05 | 0.688 | 0.713 | 0.380 | 0.570 | 0.582 |
| 0.15 | 0.489 | 0.525 | 0.203 | 0.302 | 0.363 |
|  |  | 0.538 |  |  | 0.304 |
| 0.25 | 0.348 | 0.400 | 0.121 | 0.162 | 0.233 |
|  |  | 0.401 |  |  | 0.223 |

**Table 2.** A comparison between the self–diffusivity and collective mobilities of 16 spheres, in random configurations, in a periodic domain. The diffusivity $D_T/D_o$ of one sphere in a periodic domain is based on the analytical results of Zick & Homsy.[35]

To emphasize the influence of the number of spheres, another set of simulations, with $N_P = 27$ spheres of equal diameters in a cubic periodic box, was done. $N = 40$ was used. Volume fractions of $\phi = 0.05$, 0.15 and 0.25 were considered. The spheres were arranged in a simple cubic configuration. Thus, the configuration was identical to that for a single sphere in a periodic box. The difference in the diffusivities for the two cases at a given volume fraction can then be attributed to the effect of the number of spheres. The FIMAT Stokes' problem was solved for 450 realizations at each volume fraction.

Table (3) lists the values of $D_s$ obtained from FIMAT Stokes simulations. It is seen that, as expected, the diffusivity is much larger for 27 spheres compared to a single sphere. The computed values are also compared with those by Ladd.[36] The overall trends are in reasonable agreement. The values attributed to Ladd[36] and listed in table (3), were obtained from a curve fit to his data. He did not solve for 27 spheres and none of his numbers of spheres could be arranged in a cubic configuration. As expected, the diffusivity for 27 spheres (table 3) is also larger than that for 16 spheres (table 2) at same volume fractions.

It should also be noted that the values by Ladd[36] are based on random configurations. Table (3) also lists one case for 27 random spheres at $\phi = 0.05$.

| $\phi$ | $D_T/D_o$ | $D_s/D_o$ | $D_s/D_o$ |
|---|---|---|---|
| | One sphere | FIMAT | Ladd (1990) |
| 0.05 | 0.380 | 0.758 | 0.718 |
| 0.05 (Random) | 0.380 | 0.744 | 0.718 |
| 0.15 | 0.203 | 0.612 | 0.521 |
| 0.25 | 0.121 | 0.484 | 0.376 |

**Table 3.** Self–diffusivity of 27 spheres in a periodic domain. FIMAT results are based on a simple cubic arrangement of spheres. The data due to Ladd[36] are for random configurations. The diffusivity of one sphere in a periodic domain is based on the analytical results of Zick & Homsy.[35]

### 4. Unsteady FIMAT simulations

In this section unsteady simulations of the fluctuating fluid–particle problem are reported. These computations will be referred to as the unsteady FIMAT dynamics simulations. In section 2.3, the time discrete approximation of the fluctuating hydrodynamic equations was presented. In this section an implicit fractional time step type scheme for the unsteady FIMAT simulations is presented.

A periodic computational domain $\Omega$ is considered in which a particle occupies domain P. One spherical particle is considered in this work. The fluid and the sphere densities are the same. The DLM approach for particulate flows is used with a fractional time stepping scheme.[20,39,40] The algorithm is given below:

1. <u>Solve the momentum equation:</u> Given $\mathbf{u}^n$ in $\Omega$, solve the following equation to get $\hat{\mathbf{u}}$ in $\Omega$:

$$\frac{(\hat{\mathbf{u}} - \mathbf{u}^n)}{\Delta t^*} = \nabla^2 \hat{\mathbf{u}} + \nabla \cdot (\Delta \widetilde{\mathbf{W}}), \qquad (30)$$

where $\hat{\mathbf{u}}$ is an intermediate velocity and the entire domain is assumed to be a fluid such that the particle domain moves rigidly due to a rigidity constraint (step 3 below). All the variables are scaled according to equation (8). $\Delta t^* = \Delta t / T_\mu$ (see equations 8 & 9) is the dimensionless time step scaled by the viscous time scale $T_\mu$. The properties of $\Delta \widetilde{\mathbf{W}}$ are as given in equation (9). As discussed after equation (9), the differential symbols for spatial derivatives have been used simply for the convenience of presentation. They should be considered to imply the corresponding central difference discrete operators in equation (30) and in the remaining steps of the algorithm presented below. A staggered grid is considered (see section 2.2). The equation for each velocity component is solved by using an FFT solver (FISHPAK).

2. <u>Projection on to divergence free velocity:</u> Given $\hat{\mathbf{u}}$ in $\Omega$, solve for $p^{n+1}$ and $\widetilde{\mathbf{u}}$ in $\Omega$:

$$\left.\begin{array}{l} \dfrac{(\widetilde{\mathbf{u}} - \hat{\mathbf{u}})}{\Delta t^*} = -\nabla p^{n+1}, \quad \text{where} \\[2ex] \nabla \cdot \widetilde{\mathbf{u}} = 0 \Rightarrow \nabla^2 p^{n+1} = \dfrac{1}{\Delta t^*} \nabla \cdot \hat{\mathbf{u}}. \end{array}\right\} \quad (31)$$

The equation for pressure is solved first by using an FFT solver (FISHPAK). Using the solution for $p^{n+1}$, the first of equation (31) is used to solve for $\widetilde{\mathbf{u}}$.

3. <u>Projection on to rigid motion in P:</u> Given $\widetilde{\mathbf{u}}$ in $\Omega$, set $\widetilde{\mathbf{u}} = \mathbf{u}^{n+1}$ in $\Omega/\overline{P(t)}$ and project $\widetilde{\mathbf{u}}$ in P onto rigid velocity to get $\mathbf{u}^{n+1}$ in P. $\mathbf{u}^{n+1}$ in P is calculated as follows:[20,39,40]

$$\left.\begin{array}{l} \mathbf{u}^{n+1} = \mathbf{U}^{n+1} + \boldsymbol{\omega}^{n+1} \times \mathbf{r}, \quad \text{where} \\[2ex] \dfrac{1}{6}\pi d^3 \mathbf{U}^{n+1} = \int_P \widetilde{\mathbf{u}} d\mathbf{x}, \quad \text{and} \quad \dfrac{1}{60}\pi d^5 \mathbf{I}\boldsymbol{\omega}^{n+1} = \int_P \mathbf{r} \times \widetilde{\mathbf{u}} d\mathbf{x}, \end{array}\right\} \quad (32)$$

where $\mathbf{I}$ is the identity matrix. The rigid motion constraint gives a force $\mathbf{F}^{n+1}$ in the particle domain. $\mathbf{F}^{n+1}$ in P is given by

$$\mathbf{F}^{n+1} = \frac{\left(\mathbf{u}^{n+1} - \widetilde{\mathbf{u}}\right)}{\Delta t^*}, \quad (33)$$

where the scale for $\mathbf{F}^{n+1}$ is $\mu V / L^2$.

All the variables are non–dimensional in equations (30)–(33). With the new values of $\mathbf{u}^{n+1}$, go back to step 1 to calculate the next time step. The sphere or the fluid mesoparticle (control volumes) locations are not updated because their motion is negligible (see the discussion in section 2.3). Equation (30)–(33) can be added to give the following set of equations

$$\left.\begin{array}{l}\dfrac{(\mathbf{u}^{n+1} - \mathbf{u}^{n})}{\Delta t^*} = -\nabla p^{n+1} + \nabla^2 \hat{\mathbf{u}} + \mathbf{F}^{n+1} + \nabla \cdot \left(\Delta \widetilde{\mathbf{W}}\right) \text{ in } \Omega, \\ \nabla \cdot \widetilde{\mathbf{u}} = 0 \text{ in } \Omega, \\ \nabla \cdot \mathbf{D}[\mathbf{u}^{n+1}] = \mathbf{0} \text{ in P and } \left(\mathbf{D}[\mathbf{u}^{n+1}]\right) \cdot \mathbf{n} = \mathbf{0} \text{ on } \partial \text{P}, \\ \left\langle \Delta \widetilde{W}_{ij} \right\rangle = 0, \quad \left\langle \Delta \widetilde{W}_{ik} \Delta \widetilde{W}_{lm} \right\rangle = \left(\delta_{il}\delta_{km} + \delta_{im}\delta_{kl}\right), \end{array}\right\} \quad (34)$$

Ensemble averaging of equation (34), which is linear, leads to a deterministic fractional time stepping scheme, which is first order with respect to time.[20] Thus, the discrete stochastic equations have a weak convergence of order 1 (equation 11).

To verify the approach, unsteady calculations were done for a sphere in a periodic box at volume fraction $\phi = 0.008$ (i.e., $d/\Delta h = 10$). There were 40 (= $L/\Delta h$) control volumes in each direction. The flow was driven only by random stresses. New sets of independent random stresses were computed at each time step as given in equation (34). Each sample trajectory was solved for up to 10000 time steps (each time step took around 8s real time). Several trajectories were solved.

The initial velocity field was set as follows. Random velocities were chosen in the control volumes such that their variance was as per the Maxwell–Boltzmann distribution. This random velocity field was then projected on to a divergence free velocity field in the entire domain and a rigid motion in the sphere domain. The resultant velocity field was used as the initial condition.

The velocity autocorrelation, based on sphere velocities, was computed by ensemble averaging. Its normalized value was compared with the analytical results of Hauge & Martin–Löf.[10] This is discussed below. A similar comparison was done by Ladd[5,6] to verify the Lattice–Boltzmann method for Brownian particles.

The translational velocity autocorrelation is given by $C_U(t) = \langle \mathbf{U}(0) \cdot \mathbf{U}(t) \rangle / 3$, where the variables are dimensional. Hauge & Martin–Löf[10] gave an analytical expression for $C_U(t)$

$$C_U(t) = \frac{k_B T A B}{\pi} \int_0^\infty dx\, e^{-x|t|} \frac{\sqrt{x}}{\left[A - \left(\mathrm{M} + AB^2/9\right)x\right]^2 + A^2 B^2 x}, \quad (35)$$

where $A = 3\pi\mu\,d$, $B = d/2\sqrt{\mu}$, M is the mass of the sphere and the variables are dimensional. Their governing equations were the same as those used in this work where they solve the linearized (i.e., neglecting the convection term) fluctuating hydrodynamic equations coupled with the equation of motion of a single sphere in an infinite domain. Equation (35) is valid even when the fluid and sphere densities are not the same. Hauge & Martin–Löf[10] assumed that the sphere positions are fixed, similar to the assumption in this work (section 2.3). It is seen from equation (35) that $C_U(t) \sim (2k_B T/3\rho)(4\pi\mu|t|/\rho)^{-3/2}$ at long times. This is the algebraic tail ($t^{-3/2}$) in the velocity autocorrelation function consistent with the molecular time autocorrelation functions. The classical Langevin equation gives an exponential decay for a single sphere in an infinite domain:

$$C_U(t) = \frac{k_B T}{M} e^{-3\pi\mu d|t|/M}. \qquad (36)$$

Hauge & Martin–Löf[10] showed, based on equations (35) and (36), that classical Langevin autocorrelation is valid only when the sphere density is much larger than the fluid density.

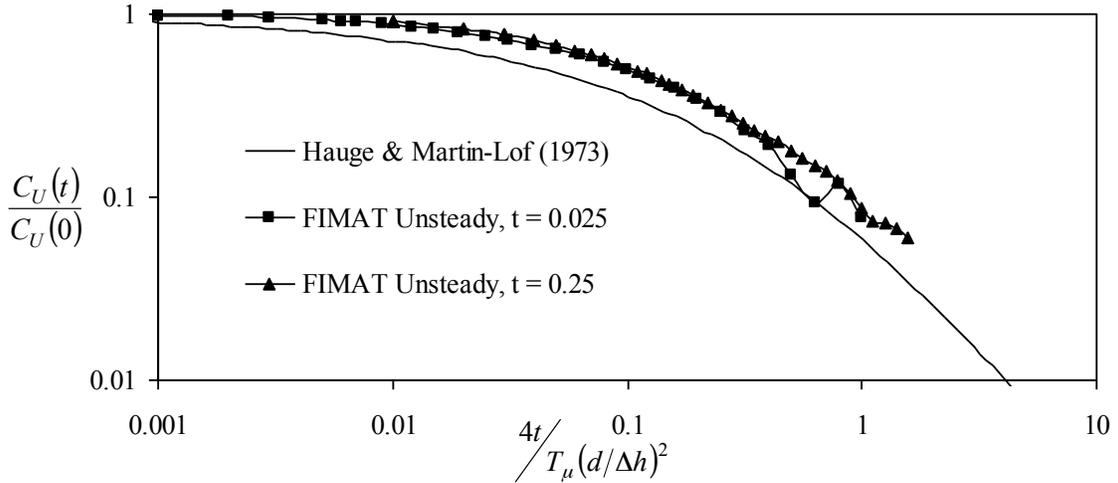

Figure 6. Plot of the normalized velocity autocorrelation vs. non–dimensional time. The fluid and sphere densities are the same.

In this work, we considered the sphere density equal to that of the fluid. The numerical results should be compared with equation (35). Figure (6) shows a plot of the normalized translational velocity autocorrelation as a function of $4t/T_\mu (d/\Delta h)^2$. The numerical values compare well with the analytical results of Hauge & Martin–Löf.[10] The comparison, between the analytical results of Hauge & Martin–Löf[10] for an infinite domain and the numerical results for a periodic domain, is found to be reasonable because the velocity autocorrelations are computed for relatively short times for a low volume fraction. The periodic boundary condition has negligible effect. This is also discussed by Ladd.[5,6]

The analytical value of $C_U(0)/(k_B T/M)$ is 2/3 due to the incompressibility constraint (see discussion by Hauge & Martin–Löf[10]). The numerical values were 0.71 with $\Delta t^* = 0.025$ and 0.88 with $\Delta t^* = 0.25$.

The angular velocity autocorrelation is defined by $C_\omega(t) = \langle \omega(0) \cdot \omega(t) \rangle / 3$. Figure (7) shows a plot of the normalized angular velocity autocorrelation as a function of $4t / T_\mu (d/\Delta h)^2$. The numerical values are compared with the analytical results of Hauge & Martin–Löf.[10] The agreement is reasonable. The numerical value of $C_\omega(0)/(k_B T/I_p)$ was, equal to 1.07 with $\Delta t^* = 0.025$ and 1.23 with $\Delta t^* = 0.25$. The analytical value is 1. Note that the numerical data in figures (6) and (7) show some fluctuations at later times due to lack of sufficient ensemble in the averaging.

Improving of the accuracy of fractional time stepping schemes is feasible,[8] and will be considered in the future.

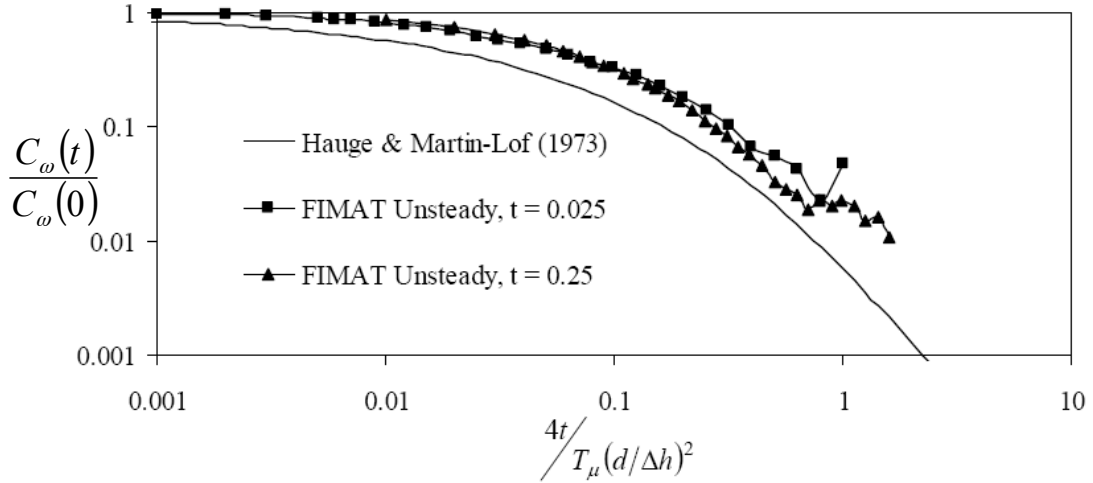

Figure 7. Plot of the normalized velocity autocorrelation vs. non–dimensional time. The fluid and sphere densities are the same.

## 5. Conclusion

A FRS scheme, named FIMAT dynamics, for the Brownian motion of particles was presented. Thermal fluctuations were included in the fluid equations via random stress terms. Solving the fluctuating hydrodynamic equations coupled with the particle equations of motion resulted in the Brownian motion of the particles. The particles acquired random motion through the hydrodynamic force acting on its surface from the surrounding fluctuating fluid. The random stresses in the fluid equations were easy to compute.

Two types of problems were considered – Stokes' problem in the long time dissipative limit and the unsteady problem for the short time behavior. Stokes' problem for a single fluid was solved first and the results were compared to the analytical solution. The agreement was found to be good.

Next, the fluid–particle problem in the long time dissipative limit was considered. We considered particles at high concentrations, particles of non–spherical shapes and many particles. The rotational diffusion of the particles is also considered here. The method was found to give good results for the Brownian motion of the particles.

Unsteady simulations, for the short time behavior of the Brownian particles, were performed by using a fractional time stepping scheme. The translational and rotational velocity autocorrelation functions, of a single sphere, were computed. The numerical simulations gave an algebraic velocity autocorrelation function, in agreement with the analytical results of Hauge & Martin–Löf.[10] This is consistent with the molecular time autocorrelation functions.

This approach can be easily incorporated into existing fluid flow solvers based on the Navier–Stokes equations. The method can be potentially extended to fluids with varying properties and temperatures. Extension of the method to droplets and elastic bodies immersed in fluids should be considered in the future.

**Acknowledgments**

This work was supported by National Science Foundation through the CAREER grant CTS–0134546. We thank Prof. Tony Ladd for many stimulating and helpful discussions.

**References**

1. Ermak, D.L. and J.A. Mccammon, *Brownian Dynamics with Hydrodynamic Interactions.* Journal of Chemical Physics, 1978. **69**(4): p. 1352-1360.
2. Brady, J.F. and G. Bossis, *Stokesian Dynamics.* Annual Review of Fluid Mechanics, 1988. **20**: p. 111-157.
3. Hoogerbrugge, P.J. and J.M.V.A. Koelman, *Simulating Microscopic Hydrodynamic Phenomena with Dissipative Particle Dynamics.* Europhysics Letters, 1992. **19**(3): p. 155-160.
4. Lei, H., B. Caswell, and G.E. Karniadakis, *Direct construction of mesoscopic models from microscopic simulations.* Physical Review E, 2010. **81**(2): p. Artn: 026704.
5. Ladd, A.J.C., *Numerical Simulations of Particulate Suspensions Via a Discretized Boltzmann-Equation .1. Theoretical Foundation.* Journal of Fluid Mechanics, 1994. **271**: p. 285-309.
6. Ladd, A.J.C., *Numerical Simulations of Particulate Suspensions Via a Discretized Boltzmann-Equation .2. Numerical Results.* Journal of Fluid Mechanics, 1994. **271**: p. 311-339.
7. Landau, L.D. and E.M. Lifshitz, *Fluid Mechanics*. 1959, London: Pergamon Press.
8. Bell, J.B., A.L. Garcia, and S.A. Williams, *Numerical methods for the stochastic Landau-Lifshitz Navier-Stokes equations.* Physical Review E, 2007. **76**(1): p. -.
9. Serrano, M. and P. Espanol, *Thermodynamically consistent mesoscopic fluid particle model.* Physical Review E, 2001. **6404**(4): p. 046115.
10. Hauge, E.H. and A. Martin-Lof, *Fluctuating hydrodynamics and Brownian motion.* Journal of Statistical Physics, 1973. **7**: p. 259-281.
11. Zwanzig, R., *Hydrodynamic Fluctuations and Stokes Law Friction.* Journal of Research of the National Bureau of Standards Section B-Mathematics and Mathematical, 1964. **B 68**(4): p. 143-&.
12. Chen, Y., N. Sharma, and N.A. Patankar, *Fluctuating immersed material (FIMAT) dynamics for the direct simulation of the Brownian motion of particles*, in *IUTAM Symposium on Computational Approaches to Multiphase Flow*, S. Balachandar and A. Prosperetti, Editors. 2004, Springer: Dordrecht. p. 119-129.
13. Patankar, N.A., *Direct numerical simulation of moving charged, flexible bodies with thermal fluctuations.* Iccn 2002: International Conference on Computational Nanoscience and Nanotechnology, ed. M. Laudon and B. Romanowicz. 2002, Cambridge: Computational Publications. 93-96.
14. Sharma, N. and N.A. Patankar, *Direct numerical simulation of the Brownian motion of particles by using fluctuating hydrodynamic equations.* Journal of Computational Physics, 2004. **201**(2): p. 466-486.
15. Atzberger, P.J., P.R. Kramer, and C.S. Peskin, *A stochastic immersed boundary method for fluid-structure dynamics at microscopic length scales.* Journal of Computational Physics, 2007. **224**(2): p. 1255-1292.
16. Donev, A., J.B. Bell, A.L. Garcia, and B.J. Alder, *A Hybrid Particle-Continuum Method for Hydrodynamics of Complex Fluids.* Multiscale Modeling & Simulation, 2010. **8**(3): p. 871-911.
17. He, P. and R. Qiao, *Self-consistent fluctuating hydrodynamics simulations of thermal transport in nanoparticle suspensions.* Journal of Applied Physics, 2008. **103**(9).